# HIGH-INTENSITY SINGLE-BUNCH ELECTRON BEAM GENERATION WITH THE 38 MeV L-BAND LINAC IN OSAKA UNIVERSITY

S. Okuda, T. Yamamoto, S. Suemine, G. Isoyama
ISIR, Osaka University, 8-1 Mihogaoka, Ibaraki, Osaka 567-0047, Japan

*Abstract*

The characteristics of the single-bunch electron beam of the 38 MeV L-band electron linac at the Institute of Scientific and Industrial Research in Osaka University have been investigated. Recently, the injector of the linac has been improved to increase the peak beam current for injection. The maximum charge of electrons in a bunch of the accelerated beam is 91 nC at an energy of 27 MeV. The energy spread is 0.8-2.5% FWHM. The bunch length measured with a streak camera is 20-30 ps FWHM. The normalized rms emittance of the beam is 140 $\pi$mm mrad at a charge of 63 nC/bunch. The possibility for obtaining beams at higher charges and the applications of the beams to generating high-power radiation are discussed.

## 1 INTRODUCTION

The 38 MeV L-band electron linac (1300 MHz) at the Institute of Scientific and Industrial Research (ISIR) in Osaka University was constructed in 1978 for generating high-intensity single-bunch beams. After the improvement of a subharmonic-prebuncher (SHPB) system the charge of electrons in the beam increased to 67 nC/bunch in maximum [1].

The main application researches using the linac are the analyses of ultra-fast phenomena induced in matters by the irradiation with the beam in a short period, and the generation of high-intensity radiation from the beams such as the self-amplified spontaneous emission with a wiggler [2] and the coherent synchrotron or transition radiation [3].

In order to increase the electron charge in the beam a newly developed high-current electron gun was installed in the linac and the locations of the magnetic lenses were optimized, which improved the characteristics of the beams for injection.

High-intensity electron beams were generated at Argonne National Laboratory (ANL) at charges up to 100 nC/bunch for the development of a wakefield accelerator [4]. They used a photo-cathode gun system.

In the present work the characteristics of the single-bunch beams of the ISIR linac in which a new injector system was installed, were investigated.

## 2 COMPONENTS OF THE ISIR LINAC

The components of the ISIR linac are schematically shown in Fig. 1. A thermionic triode gun [5] has been newly developed for the linac. The cathode-grid assembly YU-156 (EIMAC) is installed in the gun. The area of the cathode is 3.0 cm$^2$. The grid pulser of the gun has been developed by using an avalanche-type pulser. A pulsed electron beam is injected from the gun at a pulse width of 5 ns FWHM. The maximum peak beam current is 30 A at an anode voltage of 100 kV. After the installation of the gun in the linac the quality and the pulse characteristics of the beam for injection have been much improved. According to the results for the calculation of the trajectories of electrons from the gun, the locations of the two magnetic lenses between the gun and the first SHPB have been optimized.

The linac is equipped with three SHPBs (two at an rf frequency of 108 MHz and one at 216 MHz). The SHPB has a coaxial single-gap cavity. Pulsed rf at a 20 $\mu$s duration is supplied to the SHPB by three 20 kW rf amplifiers independently. To generate a single-bunch beam the pulse length of the electron beam injected from the gun to the first SHPB is 5 ns in maximum.

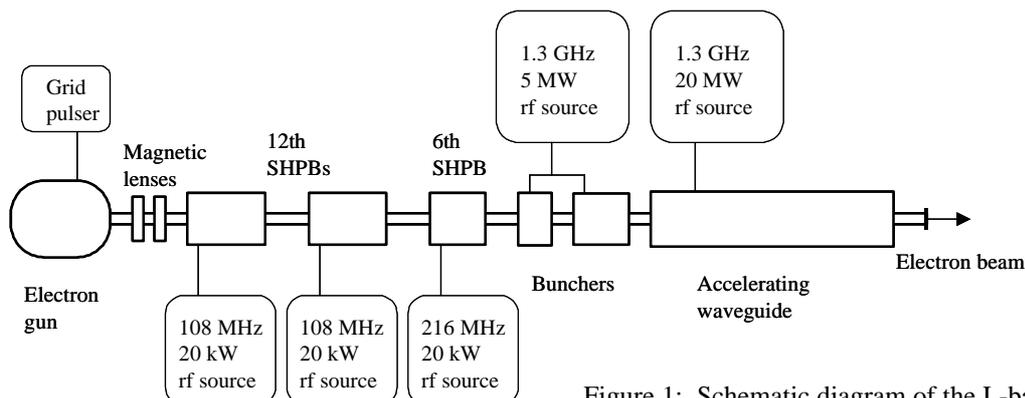

Figure 1: Schematic diagram of the L-band linac at ISIR.

Two bunchers are driven by a 5 MW L-band klystron. An accelerating waveguide 3 m long is driven by a 20 MW L-band klystron. The lengths of the pulsed rf of these klystrons on the flattop are 3.2 and 3.9 µs, respectively. The maximum beam energy is 38 MeV in the case with no beam loading. The details of the operational conditions of the rf components are described in ref. 6.

While the accelerator system is optimized to generate the single-bunch beam, multibunch beams having a macropulse length of 5 ns to 2 µs are also generated under the different operational conditions of the SHPBs. The beams are being applied to a variety of basic researches.

## 3 EXPERIMENTAL

The present experiments were performed with electron beams at an energy of 27 MeV. The energy spectrum of the beam was measured with an energy analyser composed of a bending magnet and a Faraday cup. In the measurement the energy resolution was about 0.1%. The operational conditions of the linac components were determined as to obtain the smallest energy spread. The charge of electrons in the beam was measured by using an Al block as a beam dump, placed near a beam output window in the air atmosphere. The influence of the secondary electrons emitted from the block was negligible. The error in the measurement was estimated to be below 5%. The normalized rms emittance of the electron beam was measured with a method using quadrupole magnets and a beam profile monitor. The emittance of the beam was given by averaging the values obtained for the vertical and the horizontal directions. The values of emittance for these two directions well agreed.

A time profile of the single-bunch beam was measured with a streak camera. In the measurement the Cherenkov light emitted from the beam passing in air was observed. The time resolution in the measurement was 3 ps.

## 4 RESULTS AND DISCUSSION

Figure 2 shows the energy spectrum of the single-bunch beam measured at a charge of 22 nC/bunch. This is a typical one for a wide range of beam intensity. The shape is determined by the intensity of electric field in the rf cavities, which depends on the rf phase, the beam loading effects and the wakefield induced by the electron bunch. Figure 3 shows the electron charge dependence of the energy spread. The maximum charge of the single-bunch beam obtained in the present experiments is 91 nC/bunch. The energy spread is approximately 1 % FWHM at charges below about 40 nC, and at the higher charges it increases with the charge. In the previous work the spectrum narrowing due to the wakefield from the bunch in the waveguide was observed at charges of about 30 nC [7]. This phenomenon is not apparent in Fig. 3.

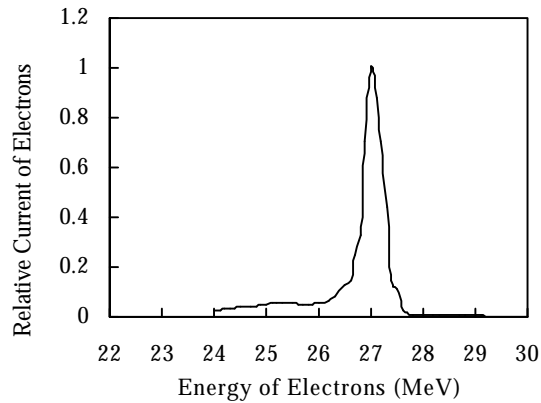

Figure 2: Typical energy spectrum of the single-bunch beam at an electron charge of 22 nC/bunch.

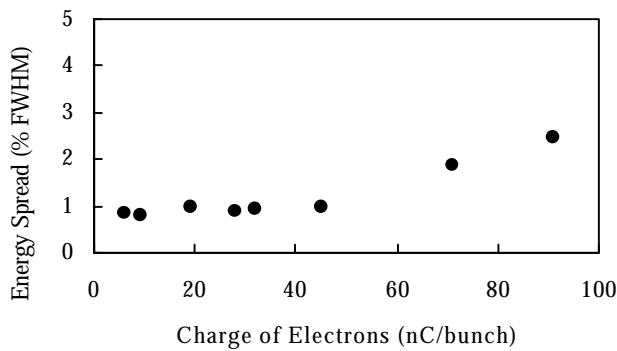

Figure 3: Electron charge dependence of the energy spread of the single-bunch beam.

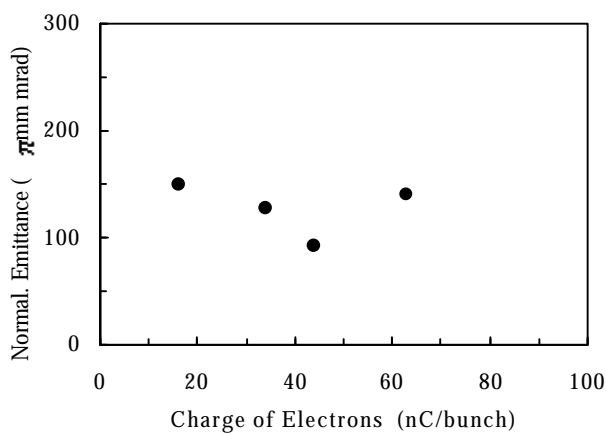

Figure 4: Electron charge dependence of the normalized rms emittance of the single-bunch beam, given by being averaged for vertical and horizontal directions.

Figure 4 shows the electron charge dependence of the normalized rms emittance of the beam. While the number of data is limited, the emittance seems to be in a range of 100-150 $\pi$mm mrad. These values are considerably low compared to those at ANL. This can be attributed to the difference in the injector system.

The typical time profile of the electron bunch measured with a streak camera is shown in Fig. 5. The electron charge is 73 nC/bunch. In most cases the bunch has a triangular profile as shown in this figure, and the bunch length is 20-30 ps FWHM. It is expected that the bunch is effectively compressed by using a chicane-type bunch compressor [8].

The results for the present work show that an extremely high electron charge in the single-bunch beam has been achieved at considerably low emittance. For the further increase of the charge the grid pulser of the gun is being improved to make the pulse shape of the beam from the gun nearly rectangular.

The single-bunch beams are being used for exciting matters strongly and for generating high-intensity radiation. The radiation is now being applied as new far-infrared light sources to spectroscopy. For application researches of the radiation, the experiments for developing a new method for high-gradient particle acceleration are being prepared.

## 5 CONCLUSIONS

The characteristics of the electron beams for injection were improved by the installation of the new injector system in the ISIR L-band linac. The single-bunch beams were generated at an extremely high intensity and at relatively good beam quality. The maximum charge of electrons was 91 nC/bunch at an energy of 27 MeV. The bunch lengths were 20-30 ps FWHM. The charge will increase by improving the pulse characteristics of the grid pulser of the gun. The high-power radiation is being applied to new research fields.

## ACKNOWLEDGEMENTS

The authors would like to thank Dr. T. Kozawa, Mr. T. Igo and Dr. R. Kato for the technical help in measuring the bunch shape and the beam emittance.

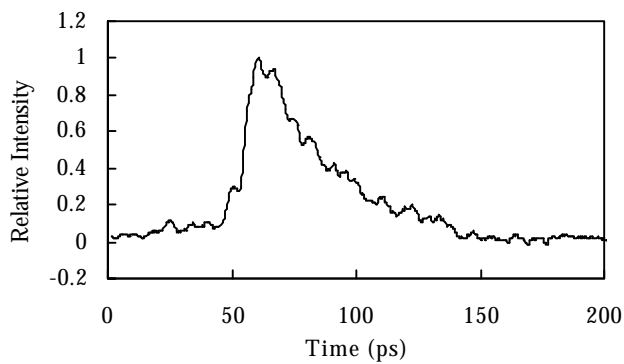

Figure 5: Time profile of the single-bunch beam measured with a streak camera at an electron charge of 73 nC/bunch.